\documentclass[runningheads]{llncs}
\usepackage{graphicx}
\usepackage{import}

\usepackage[nogroupskip=true, style=super, nonumberlist]{glossaries}

\newacronym{IVA}{IVA}{Intelligent Video Analytics}
\newacronym{RIVA}{RIVA}{Real-time IVA}
\newacronym{BIVA}{BIVA}{Batch IVA}
\newacronym{VAS}{VAS}{Video Analytics System}
\newacronym{EVAS}{EVAS}{Embedded Video Analytics System}
\newacronym{OVAS}{OVAS}{On-site Video Analytics System}
\newacronym{CVAS}{CVAS}{Cloud-based Video Analytics System}
\newacronym{FVAS}{FVAS}{Fog-based Video Analytics System}
\newacronym{IR}{IR}{Intermediate Results}
\newacronym{VSDS}{VSDS}{Video Stream Data Source}
\newacronym{IVAS}{IVAS}{IVA services}
\newacronym{SIAT}{L-CVAS}{Lambda CVAS}
\newacronym{aaS}{aaS}{as-a-Service}
\newacronym{IaaS}{IaaS}{Infrastructure-as-a-Service}
\newacronym{PaaS}{PaaS}{Platform-as-a-Service}
\newacronym{SaaS}{SaaS}{Software-as-a-Service}
\newacronym{IVAaaS}{IVAaaS}{IVA-as-a-Service}
\newacronym{IVAAaaS}{IVAAaaS}{IVA-Algorithm-as-a-Service}
\newacronym{CBVR}{CBVR}{Context-Based Video Retrieval}
\newacronym{ITS}{ITS}{Intelligent Transportation System}
\newacronym{ML}{ML}{Machine Learning}
\newacronym{SVM}{SVM}{Support Vector Machine}
\newacronym{VBDCL}{VBDCL}{Video Big Data Curation Layer}
\newacronym{VBDPL}{VBDPL}{Video Big Data Processing Layer}
\newacronym{VBDML}{VBDML}{Video Big Data Mining Layer}
\newacronym{KCL}{KCL}{Knowledge Curation Layer }
\newacronym{WSL}{WSL}{Web Service Layer}
\newacronym{RVSAS}{RVSAS}{Real-time Video Stream Acquisition and Synchronization}
\newacronym{DPDS}{DPDS}{Distributed Persistent Data Store}
\newacronym{DMBM}{DMBM}{Distributed Message Broker Manager}
\newacronym{VSAS}{VSP}{Video Stream Producer}
\newacronym{IRM}{IRM}{Intermediate Results Manager}
\newacronym{VSCS}{VSC}{Video Stream Consumer}
\newacronym{LVSM}{LVSM}{Lifelong Video Stream Monitor}
\newacronym{VSAC}{VSAC}{Video Stream Analytics Consumer}
\newacronym{ISDDS}{ISDDS}{Immediate Structured Distributed Data Store}
\newacronym{UPDDS}{UPDDS}{Unstructured Persistent Distributed Data Store}
\newacronym{DFS}{DFS}{Distributed File System}
\newacronym{DBDS}{DBDS}{Distributed Big Datastore}
\newacronym{HDFS}{HDFS}{Hadoop File System}
\newacronym{LSH}{LSH}{Locality-sensitive Hashing}
\newacronym{MCA}{MCA}{Multiple Correspondence Analysis}
\newacronym{AFS}{AFS}{Adaptive Feature Scaling}
\newacronym{SIFT}{SIFT}{Scale Invariant Feature Transform}
\newacronym{LBP}{LBP}{Local Binary Pattern}
\newacronym{HOG}{HOG}{Histogram of Oriented Gradients}
\newacronym{CNN}{CNN}{Convolutional Neural Network}
\newacronym{DAG}{DAG}{Directed Acyclic Graph}
\newacronym{CCDG}{CCDG}{Controlled Cyclic Dependency Graph}
\newacronym{RTSP}{RTSP}{Real-time Stream Protocol}
\newacronym{PCA}{PCA}{Principal Component Analysis}
\newacronym{SM}{SM}{Similarity Measures}
\newacronym{LSI}{LSI}{Latent Semantic Indexing}
\newacronym{RNN}{RNN}{Recurrent neural network}
\newacronym{LSTM}{LSTM}{Long Short-Term Memory}
\newacronym{FVSA}{FVSA}{Fused Video Surveillance Architecture}
\newacronym{C2C}{C2C}{Customer-to-Customer}
\newacronym{B2B}{B2B}{Business-to-Business}
\newacronym{B2C}{B2C}{Business-to-Customer}
\newacronym{QoS}{QoS}{Quality of Service}
\newacronym{ACID}{ACID}{Atomicity, Consistency, Isolation, Durability}
\newacronym{RDBMS}{RDBMS}{Relational Database Management System}
\newacronym{CAP}{CAP}{Consistency, Availability, Partition Tolerance}
\newacronym{CRUD}{CRUD}{create, read, update, and delete}
\newacronym{AI}{AI}{Artificial Intelligence}
\newacronym{IoT}{IoT}{Internet of Things}

\makeglossaries

\begin{document}
%
\title{Video Big Data Analytics in the Cloud: Research Issues and Challenges
	\thanks{This work was supported by the Institute for Information and Communications Technology Promotion Grant through the Korea Government (MSIT) under Grant R7120-17-1007 (SIAT CCTV Cloud Platform).}}
%
%
\author{Aftab Alam\inst{1}
	\and
Shah Khalid\inst{2}
\and
Muhammad Numan Khan\inst{1}
\and
Tariq Habib Afridi \inst{1}
\and
Irfan Ullah\inst{1}
\and
Young-Koo~Lee\inst{1}}
%
\authorrunning{A. Alam et al.}
%
\institute{Department of Computer Science and Engineering, Kyung Hee University (Global Campus), Yongin 1732, South Korea \\
\email{\{aftab,numan,afridi,irfan,yklee\}@khu.ac.kr}
\and
School of Computer Science and Communication Engineering, Jiangsu University \\
\email{\{shahkhalid\}@ujs.edu.cn}}
\maketitle              
\begin{abstract}
On the rise of distributed computing technologies, video big data analytics in the cloud have attracted researchers and practitioners' attention. The current technology and market trends demand an efficient framework for video big data analytics. However, the current work is too limited to provide an architecture on video big data analytics in the cloud, including managing and analyzing video big data, the challenges, and opportunities. This study proposes a service-oriented layered reference architecture for intelligent video big data analytics in the cloud. Finally, we identify and articulate several open research issues and challenges, which have been raised by the deployment of big data technologies in the cloud for video big data analytics. This paper provides the research studies and technologies advancing the video analyses in the era of big data and cloud computing. This is the first study that presents the generalized view of the video big data analytics in the cloud to the best of our knowledge.


\end{abstract}

\section{Introduction}
Videos are recorded and uploaded to the cloud on a regular base. Many sources include CCTV, smartphones, drones, satellites, etc. They are actively contributing to video generation, leading to the evolution of video analytics and management systems.  Video management and services providers such as Facebook \cite{Facebook2020}, and YouTube \cite{YouTube2019}, are considered as valuable sources of large-scale video data. Along with these, various leading industrial organizations have successfully deployed video management and analytics platforms that provide more bandwidth and high-resolution cameras collecting videos at scale and has become one of the latest trends in the video surveillance industry. For example, more than 400 hours of videos are uploaded in a minute on Youtube \cite{pouyanfar2018multimedia}, and more than one hundred and seventy million video surveillance cameras have been installed in china only \cite{olatunji2018dynamic}. It has been reported that the data generated by various \gls{IoT} devices will see a growth rate of 28.7\% over the period 2018-2025, where surveillance videos are the majority shareholder \cite{idg2019}.

Such an enormous video data is considered “big data” because various sources generate a large volume of video data at high velocity that holds high Value. Even though 65\% of the big data shares hold by surveillance videos are monitored, but still, a significant portion of video data has been failed to notice\cite{huang2014surveillance}. That neglected data contain valuable information directly related to real-world situations. Video data provide information about interactions, behaviors, and patterns, whether its traffic or human patterns. However, handling such a large amount of complex video data is not worthwhile utilizing traditional data analytical approaches. Therefore, more comprehensive and sophisticated solutions are required to manage and analyses such unstructured video big data.

Due to the data-intensive and resources hungry nature of large-scale video data processing, extracting the video's insights is a challenging task. A considerable video data video poses significant challenges for video management and mining systems that require powerful machines to deal with video big data. Moreover, a flexible solution is necessary to store and mine this large volume of video data for decision making. However, large-scale video analytics becomes a reality due to the popularity of big data and cloud computing technologies.  

Cloud computing is an infrastructure for providing convenient and ubiquitous remote access to a shared pool of configurable computing resources. These resources can be managed with minimal management effort or service. Big data technologies, such as Hadoop or Spark echo system, are software platforms designed for distributed computing to process, analyze, and extract the valuable insights from large datasets in a scalable and reliable way. The cloud is preferably appropriate to offer the big data computation power required to process these large datasets. Amazon web service \cite{amazon2015amazon} and Oracle Big Data Analytics \cite{dijcks2012oracle} are some examples of big data analytics platforms. When video analytics solutions are provided in a cloud computing environment, then it is called \gls{CVAS}. Large-scale video analytics in the cloud is a multi-disciplinary area, and the next big thing in big data, which opens new research avenues for researchers and practitioners.

This work aims to conduct a comprehensive study on the status of large-scale video analytics in the cloud-computing environment while deploying video analytics techniques. First, a lambda style \cite{marz2015big} service-oriented reference architecture called \gls{SIAT} has been proposed for video big data analytics in the cloud. Then open research issues and challenges are discussed, with a focus on proposed architecture. 


\section{Proposed L-CVAS Architecture}
\label{sec:CVAS}
\begin{figure}
	\centerline{\includegraphics[width=10cm]{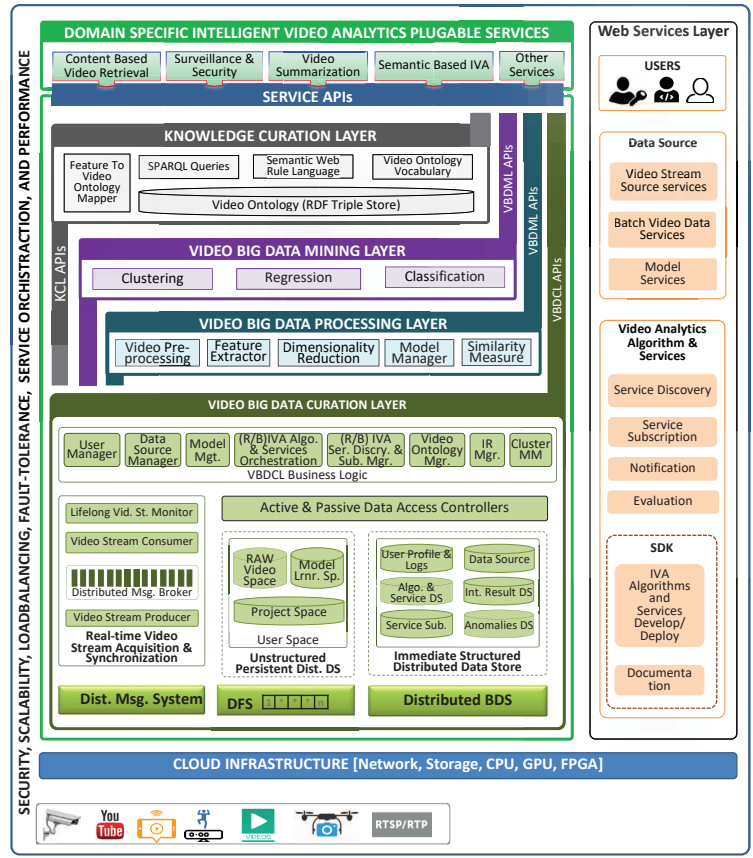}}
	\caption{A reference architecture for intelligent video analytics in the cloud.} \label{fig:IVAaaSRefArchitecture}
\end{figure}

Fig.~\ref{fig:IVAaaSRefArchitecture} presents the proposed \gls{SIAT} reference architecture for distributed \gls{IVA} in the cloud. It is composed of five layers i.e., \gls{VBDCL}, \gls{VBDPL}, \gls{VBDML}, \gls{KCL}, and \gls{WSL}.

\subsection{Video Big Data Curation Layer}
\label{Sec:VBDCL}
Effective data management is key to extract insights from the data. It is a  petascale storage architecture that can be accessed in a highly efficient and convenient manner. We design the \gls{VBDCL} to manage video big data efficiently. \gls{VBDCL}’s consists of three main components: \gls{RVSAS}, \gls{DPDS}, and \gls{VBDCL} Business Logic.

\subsubsection{Real-time Video Stream Acquisition and Synchronization}

\label{sec:RVSAS}
To handle large-scale video stream acquisition in real-time, to manage the \gls{IR}, anomalies, and the communication among \gls{RIVA} services, we design the \gls{RVSAS} component while assuming a distributed messaging system. \gls{RVSAS} provides client APIs on the top of a distributed messaging system for the proposed framework. Distributed Message Broker, e.g., Apache Kafka \cite{kreps2011kafka}, is an independent application responsible for buffering, queuing, routing, and delivering the messages to the consumers being received from the message producer. The \gls{RVSAS} component is responsible for handling and collecting real-time video streams from device-independent video data sources. Once the video stream is acquired, then it is sent temporarily to the distributed broker server. The worker system, on which an \gls{IVA} service is configured, e.g., activity recognition, reads the data from the distributed broker and process. The \gls{RVSAS} component is composed of five sub-modules, i.e., 
\gls{DMBM}, \gls{VSAS}, \gls{VSCS}, \gls{IRM}, and \gls{LVSM}.

\begin{figure}
	\centerline{\includegraphics[scale=.4]{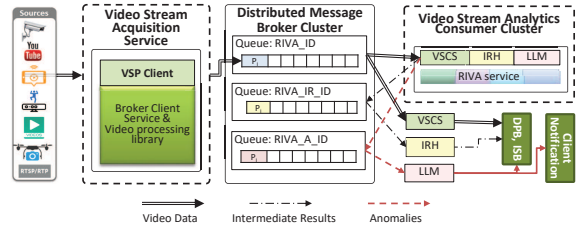}}
	\caption{Real-time Video Stream Acquisition and Synchronization} \label{fig:RVSAS}
\end{figure}

\gls{DMBM} are used to manage the queues in the distributed message broker cluster considering \gls{RIVA} services. Three types of queues, \texttt{RIVA\_ID}, \texttt{RIVA\_IR\_ID}, and \texttt{RIVA\_A\_ID} as shown in Fig.~\ref{fig:RVSAS}, are automatically generated by the \gls{DMBM} module  on the distributed message broker when a new real-time \gls{RIVA} service is created. Here \texttt{RIVA}, \texttt{ID}, \texttt{IR}, and \texttt{A} stands for \gls{RIVA} service, unique identifier of the service, \gls{IR}, and Anomalies, respectively. These queues are used to hold the actual video stream being acquired by \gls{VSAS}, \gls{IR} produced by an algorithm, and anomalies detected by the video analytics services. The \gls{VSAS} module is used to provide interfaces to video stream data sources, acquire large-scale streams from device-independent video data sources, and send it to the broker server queue, i.e., \texttt{RIVA\_ID}, in the form of compressed mini-batches. The \gls{VSCS} assists the \gls{RIVA} service to read the mini-batches of the video stream from the respective queue for analytics, as shown in Fig.~\ref{fig:RVSAS}. A \gls{RIVA} is subject to produces two types of results, i.e., intermediate results and anomalies. In this context, \gls{RIVA} utilizes \gls{VSCS} and \gls{LVSM} to sent the intermediate results and anomalies to the \texttt{RIVA\_IR\_ID}, and \texttt{RIVA\_A\_ID} queues respectively.

\subsubsection{Immediate Structured Distributed Datastore}
The \gls{ISDDS} is provided to manage large-scale structured data in the distributed environment over \gls{DBDS}. Because of the data-intensive operation and according to the other layer's requirements, technologically, a distributed big data store can be deployed such as Cassandra, HBase \cite{vora2011hadoop},  etc. The \gls{ISDDS} hosts five types of data. Role-based user logs are maintained through the User Profile and Logs meta-store. The proposed framework manages two types of video data sources through the Data Source meta-store. These are video data sources, such as IP-cameras, Kinect, body-worn cameras, etc., and batch video datasets. The former can be subscribed to \gls{RIVA} service while the latter is eligible for \gls{BIVA} services. The meta-information of these sources, along with access rights, are managed through the Data Source meta-store. Administrator and developer roles can develop, create, and deploy video analytics algorithms through the \gls{SIAT}. Similarly, different \gls{IVA} algorithms can be pipelined into an \gls{IVA} service. The management of video analytics algorithms and services is managed through Video Analytics Algorithm and Service meta-store, respectively. As stated that in \gls{IVA} pipelining environment, the output of one \gls{IVA} algorithm can be the output of another algorithm. In this context, we design a general container called \gls{IR} datastore to persist and index the output of an \gls{IVA} algorithm and services. Finally, the users are allowed to subscribe to the data sources to the \gls{IVA} services. The subscription information is maintained through the Subscription meta-store, and the anomalies are maintained through the Anomalies meta-store. 

\subsubsection{Unstructured Persistent Distributed Datastore} 
The \gls{UPDDS} component built on the top of the \gls{DFS}, such as \gls{HDFS}, that facilitates permanent and distributed big-data storage. Upon new registration, a formal User Space is created on the top of \gls{DFS}. The User Space is managed through a proper hierarchical directory structure, and special read and writes access are granted to the owner. All the directories are synchronized and mapped in the corresponding user profile logs. Under the User Space, three types of distributed directories are created, i.e., Raw Video Space, Model Space, and Project Space. Raw Video Space is used for the management of the video data. The Model Space is provided to facilitate the developers to manage the training and testing model according to the deployed \gls{IVA} algorithm.  The Project Space is provided to manage the source code of the respective developer and practitioners.

\subsubsection{Active and Passive Data Readers and Writer} 
This module gives read-write access to the underlying data securely according to the business logic of the \gls{VBDCL} Business Logic and according to the registered user access rights. This sub-module is composed of two types of readers and writers, i.e., Active and Passive Data Reader, which are used to access  \gls{ISDDS} and \gls{UPDDS}.

\subsubsection{ISBDS Business Logic}
The actual business logic is provided by the \gls{VBDCL} Business Logic, which implements seven modules. The \texttt{User Manager} module encapsulates all the user-related operations such as new user account creation, access role assignment, and session management. Through the \texttt {Data Source Manager} and \texttt{Model Manager} modules, the user can manage the \gls{VSDS}, video data uploading, and model management. The \texttt{(R/B)IVA Algorithm and Service Manager} are built to manage, develop, and deploy new \gls{IVA} algorithms and services, respectively. The former is provided \gls{aaS} to the developers, while the latter is provided \gls{aaS} to the consumers. The developer role can create and publish a new video analytics algorithm. The algorithm is then made available \gls{aaS} to other developers and can use it. Once \gls{IVA} services are created, then the users are allowed to subscribe to the streaming video data sources and batch data against the provided \gls{RIVA} and \gls{BIVA} services, respectively, using the \texttt{Service Discovery and Subscription Manager}. The \gls{IR} Manager provides a secure way of getting the \gls{IR} and maps it according to ontology. Finally, to provide the proposed system's functionality over the web, it incorporates top-notch functionality into simple unified role-based web services. The \texttt{Web Service Layer} is built on the top of \texttt{VBDCL Business Logic}. 

\subsection{Video Big Data Processing Layer}
\label{sec:VBDPL}
\gls{IVA} requires video data pruning and strong feature extraction. With such intentions, the \gls{VBDPL} layer consists of three components, i.e., Video Pre-processing, Feature Extractor, and Dimensionality Reduction.

\texttt{Video Pre-processing} component is designed to clean and remove noise from videos. It is supposed to deploy several distributed video pre-processing operations, including frame extraction, frame-resizing, frame-conversion from RGB to grayscale, shot boundary detection, segmentation, transcoding, etc. In the first step, frames are extracted from a video for processing. The number of frames to be extracted is dependent on the user objective and task-dependent. Candidate frames can be all frames, step frames (every second frame, fifth frame, etc.), or keyframes. The spatial operations highly depend on the scenario and objective. Spatial operations include frame resizing (for reducing computational complexity), corrections (brightness, contrast, histogram equalization, cropping, keyframes), mode (RGB, Grayscale, etc.), and many other operations. Segmentation is used for various purposes, such as partitioning video into semantically related chunks. 

The \texttt{Feature Extractor} extracts the features from the raw videos that can be interpreted by the \gls{ML} algorithm. In this context, several feature extraction algorithms have been introduced for video data. These feature extraction approaches can be categorized into static features of keyframes, object features, dynamic/motion feature extraction, trajectory-based features extraction, and deep learning-based feature extraction \cite{uddin2019feature}.  

The \texttt{Feature selection and dimensionality reduction} reduce the size of the features. Large sizes of feature sets are expensive in terms of time for training and/or performing classification acquired by trained classifiers. For example, \gls{PCA} and its variants are used to reduce the size of features.  During feature selection, most relevant features are selected by discarding irrelevant and weak features. Inappropriate or partially relevant features can negatively affect model performance. Therefore, only a limited set of features should be selected and used for training classifiers. This is what precisely the purpose of this component is and deploy different algorithms in this context. Similarly, some feature reduction techniques available that selects the specific set of limited features in real-time. 

\subsection{Video Big Data Mining Layer}
\label{sec:VBDML}
The \gls{VBDML} utilizes diverse types of machine-learning algorithms, i.e., supervised, semi-supervised, and unsupervised algorithms to find different type of information from the videos \cite{xie2008event}. In this context, \gls{VBDML} layer hosts three types of components, i.e., Classification, Regression, Clustering. 

The classification component provides various \gls{ML} algorithms, e.g. \gls{SVM}, Nearest Neighbors, Random Forest, Decision Tree, Naïve Bayes, etc., that identifies that a particular object in a video frame belongs to which category while using predefined classes. The Regression component includes algorithms, e.g., Linear Regression, Decision Tree Regression, Logistic Regression, and many more., which predict a continuous-valued attributed associated with objects rather than discrete values. The Clustering component encapsulates algorithms, e.g., K-Mean, spectral clustering, etc. that produces data groups depending upon the similarity of data items.

\subsubsection{Distributed Deep Learning for IVA}
\label{sec:DLDVA}
Recently, \gls{CNN} based approaches have shown performance superiority in tasks like optical character recognition \cite{borisyuk2018rosetta}, and object detection \cite{lecun2015deep}. The motive of the deep learning is to scale the training in three dimensions, i.e., size and complexity of the models \cite{dean2012large}, proportionality of the accuracy to the amount of training data \cite{hestness2017deep}, and the hardware infrastructure scalability \cite{zhang2017poseidon}. A \gls{CNN} or ConvNet is a type of neural network that can recognize visual patterns directly from the pixels of images with less preprocessing. \gls{CNN} based video classification methods have been proposed in the literature to learn features from raw pixels from both short video and still images \cite{yue2015beyond}. In the proposed \gls{SIAT} framework, both the \gls{VBDPL}, and \gls{VBDML} are capable to deploy deep-learning approaches for distributed \gls{IVA}. Since on the dawn of deep learning, various open-source architecture have been developed. Some of the well-known and state-of-the-art \gls{CNN} architectures are AlexNet \cite{krizhevsky2012imagenet}, GoogleNet \cite{43022GoogleNet2015}, VGGNet \cite{simonyan2014very}, and ResNet \cite{he2016deep}. 

\begin{figure}
	\centerline{\includegraphics[width=\textwidth]{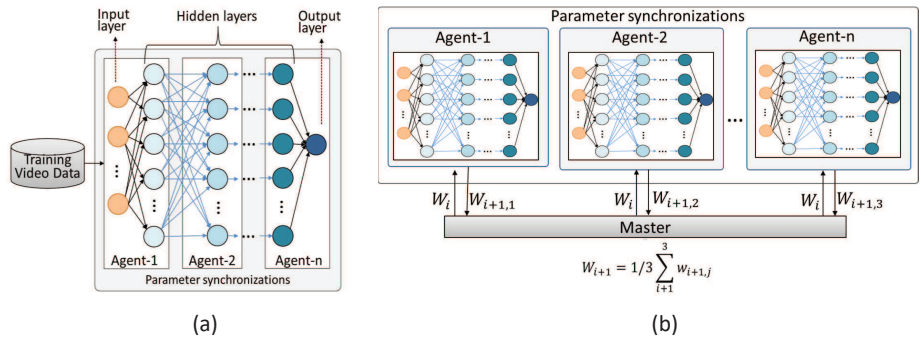}}
	\caption{(a) Scalable deep learning utilizing model distribution. (b) Scalable deep learning utilizing data distribution.} \label{fig:DLDistribution}
\end{figure}

There are two approaches for leveraging DL in distributed computing, i.e., model and data distribution \cite{mayer2019scalable}. In the former, the DL model is partitioned in logical fragments and loaded to different worker agents for training, as shown in Fig.~\ref{fig:DLDistribution}a. The training data are input to the work-agent(s) that have the input layer. In the second approach, the deep learning model is replicated to all the cluster's worker-agents, as shown in Fig.~\ref{fig:DLDistribution}b. The training dataset is partitioned into non-overlapping sub-dataset, and each sub-dataset are loaded to the different worker-agents of the cluster. Each worker-agent executes the training on its sub-dataset of training data. The model parameters are synchronized among the cluster worker-agents to updates the model parameters. The data distribution approached naturally fits in the distribute computing MapReduce paradigm \cite{deshpande2018artificial}. 

\subsubsection{Big Data Engines, ML Libraries, and IVA}
\label{sec:DistrustedProcessingEnginesAndMLLibraries}
The \gls{VBDPL}, and \gls{VBDML} are assumed to be built on the top of distributed computing engines. Hadoop MapReduce \cite{dean2010mapreduce} is a distributed programming model developed for data-intensive tasks. Apache Spark follows a similar programming model like MapReduce but extends it with Resilient Distributed Datasets (RDDs), data sharing abstraction \cite{zaharia2016apache}. Hadoop’s MapReduce operations are heavily dependent on the hard disk while Spark is based on in-memory computation, making Spark a hundred times faster than Hadoop \cite{zaharia2016apache}. Spark support interactive operations, \gls{DAG} processing, and process streaming data in the form of mini-batches in near real-time \cite{zaharia2013discretized}. Apache Spark is batch centric and treats stream processing as a special case, lacking support for cyclic operations, memory management, and windows operators. Such issues of Spark has been elegantly addressed by Apache Flink \cite{carbone2015apache}. Apache Flink treats batch processing as a special and does not use micro-batching. Similarly, Apache Storm and  Samza is another prominent solution focused on working with large data flow in real-time. 

To achieve scalability, big data techniques can be exploited by existing video analytics modules. The \gls{VBDPL} is not provided by default and needs its implementation on the top of these big data engines. However, The \gls{ML} approaches can be categorized into two classes in the context of \gls{VBDML}. One class re-implements the existing \gls{ML} task by providing a middleware layer to run them on a big data platform. This general type of middleware layer provides general primitives/operations that assists in various learning tasks. Users who want to try different \gls{ML} algorithms in the same framework can take benefits from it. In the second class, the individual video analytics and \gls{ML} algorithm are executed on a big data platform that is directly built on top of a big data engine for better scalability. Spark MLlib \cite{meng2016mllib}, Mahout \cite{owen2012mahout}, FlinKML \cite{carbone2015apache} are list of some open-source \gls{ML} packages built on the top of Hadoop, Apache Spark and Flink, respectively, that support many scalable learning algorithms. For deep learning, various open-source libraries have been develop including TensorFlow \cite{abadi2016tensorflow}, DeepLearning4J \cite{team2016deeplearning4j}, Keras \cite{chollet2015keras}, BigDL \cite{dai2018bigdl}, and PyTorch \cite{ketkar2017introduction}. 

\subsection{Knowledge Curation Layer}
\label{sec:KCL}
The \gls{KCL} layer has been proposed under \gls{CVAS} architecture, on the top of \gls{VBDML}, which map the \gls{IR} (both online and offline) into the video ontology in order to allow domain-specific semantic video and complex event analysis. The \gls{KCL} is composed of five components, i.e., Video Ontology Vocabulary, Video Ontology, Semantic Web Rules, FeatureOnto Mapper, and SPARQL queries. \texttt{Video Ontology Vocabulary} standardizes the basic terminology that governs the video ontology, such as concept, attributes, objects, relations, video temporal relation, video spatial relation, and events. \texttt{Video Ontology} is a generic semantic-based model for the representation and organization of video resources that allow the \gls{CVAS} users for contextual complex, event analysis, reasoning, search, and retrieval. \texttt{Semantic Web Rules} express domain-specific rules and logic for reasoning.  When videos are classified and tagged by the \gls{VBDML} then the respective \gls{IR} are persistent to \gls{VBDCL} and also mapped to the \texttt{Video Ontology} while using the \texttt{FeatureOnto Mapper}. Finally, SPARQL based semantic rich queries are allowed for knowledge graph, complex event reasoning, analysis, and retrieval.
\section{Research Issues, Opportunities, and Future Directions}
\label{sec:ResearchIssues}
Intelligent video big data analytics in the cloud opens new research avenues, challenges, and opportunities. This section provides in-depth detail about such research challenges (summarized in Table~\ref{tab:ResearchIssues}).

\begin{table*}[!htb]
	\centering
	\caption{Video Big Data Analytics Open Research Issues and Challenges in the Cloud.}
	\label{tab:ResearchIssues}
	\resizebox{\textwidth}{!}{%
		\begin{tabular}{|l|l|l|l|}
			\hline
			\textbf{Component} &
			\textbf{Aspect} &
			\textbf{Layer} &
			\textbf{Open Research Issues} \\ \hline
			Video Big Data &
			Volume &
			VBDCL &
			Orchestration and Optimization of \gls{IVA} Pipeline \\ \hline
			&
			
			&
			VBDCL, VBDPL, VBDML &
			Big dimensionality reduction, and indexing \\ \hline
			
			&
			
			&
			
			&
			Cleaning and compressing video big data
			\\ \hline
			&
			Velocity &
			VBDCL, VBDPL, VBDML & 
			Real-time video streams and online learning \\ \hline
			&
			Variety &
			&
			\begin{tabular}[c]{@{}l@{}}Big dimensionality reduction\\ Data modality for single \gls{IVA} goal\end{tabular} \\ \hline
			&
			Veracity &
			&
			\begin{tabular}[c]{@{}l@{}}Vide big data veracity assessment \\ Learning with unreliable data\end{tabular} \\ \hline
			&
			Value &
			&
			\begin{tabular}[c]{@{}l@{}} Understandable \gls{IVA} for decision support \end{tabular} \\ \hline
			&
			
			&
			&
			Semantic concepts extraction in distributed environment \\ \hline
			User &
			\begin{tabular}[c]{@{}l@{}}
				Developer / \\ Researcher 
			\end{tabular}	
			&
			WSL &
			\begin{tabular}[c]{@{}l@{}}Declarative \gls{IVA}\\ \gls{IVA} and distributed computing technologies abstraction.\\ Comprehensive evaluation measures for \gls{IVA} in the cloud environment \\ Visualizing video big data\end{tabular} \\ \hline
			&
			&
			VBDCL, WSL &
			\gls{IVA} algorithm, model, and services statistics maintenance and ranking \\ \hline
			&
			&
			&
			Effective price scheme for \gls{IVA} algorithm deployment, and subscription \\ \hline
			&
			&
			WSL, VBDCL &
			Model management and algorithm selection \\ \hline
			&
			Consumer &
			WSL &
			\begin{tabular}[c]{@{}l@{}}\gls{IVA} services utilization\\ Improving consumer experience based on feedback\end{tabular} \\ \hline
			&
			&
			\begin{tabular}[c]{@{}l@{}}	
				VBDCL, VBDPL, \\ VBDML, VKCL 
			\end{tabular}
			
			&
			Effective price scheme for multiple \gls{IVA} service subscription \\ \hline
			&
			Security and Privacy &
			- &
			Privacy preserving distributed IVA, security, and trust \\ \hline
			Cloud System &
			Analytics engine &
			VBDPL, VBDML &
			\begin{tabular}[c]{@{}l@{}}\gls{IVA} on video big data (general big data middleware for IVA)\\ Parameter Server optimization\end{tabular} \\ \hline
			&
			Infrastructure &
			&
			\gls{IVA} on video big data (general big data middleware for \gls{IVA}) \\ \hline
		\end{tabular}%
	}
\end{table*}

\textbf{IVA on Video Big data:} Big data analytics engines are the general-purpose engine and are not mainly designed for big video analytics. Consequently, video big data analytics is challenging over such engines and demand optimization. Almost all the engines are inherently lacking the support of elementary video data structures and processing operations. Further, such engines are also not optimized, especially for iterative \gls{IVA} and dependency among processes. Furthermore, the focus of the existing research on \gls{IVA} are velocity, volume, velocity, but the veracity and value have been overlooked. One promising direction in addressing video big data veracity is to research methods and techniques capable of accessing the credibility of video data sources so that untrustworthy video data can be filtered. Another way is to come up with novel \gls{ML} models that can make inferences with insufficient video data. Likewise, users’ assistance is required to comprehend \gls{IVA} results and the reason behind the decision to realize the value of video big data in decision support. Thus, understandable \gls{IVA} can be a significant future research area.

\textbf{IVA and Human-machine coordination:} \gls{IVA} on video big data grants a remarkable opportunity for learning with human-machine coordination for numerous reasons. \gls{IVA} on video big data in cloud demands researchers and practitioners mastering both \gls{IVA} and distributed computing technologies. Bridging both the worlds for most analysts is challenging. Especially in an educational environment, where the researcher focuses more on the understanding, configuration, and tons of parameters rather than innovation and research contribution. Thus there is a growing need to design such \gls{SIAT} that provide high-level abstractions to hide the underlying complexity. \gls{IVA} service to become commercially worthwhile and to achieve pervasive recognition, consumer lacking technical \gls{IVA} knowledge. The consumers should be able to configure, subscribe, and maintain \gls{IVA} services with comfort. In traditional \gls{IVA}, consumers are usually passive. Further, research is required to build more interactive \gls{IVA} services that assist consumers in gaining insight into video big data. 

\textbf{Orchestration and Optimization of IVA Pipeline:} The real-time and batch workflow are deeply dependent on the messaging middleware and distributed processing engines. The dynamic (R/B)\gls{IVA} service creation and multi-subscription environment demand the optimization and orchestration of the \gls{IVA} service pipeline \cite{alam2020tornado} while guarantees opportunities for further research. In the map-reduce infrastructure, a slowdown predictor can be utilized to improve the agility and timeliness of scheduling decisions. Spark and Flink can accumulate a sequence of algorithms into a single pipeline but need research to examine its behavior in dynamic service creation and subscription environment. Further, concepts from the field of query and queuing optimization can be utilized while considering messaging middleware and distributed processing engines with the aim of orchestrating and optimization of \gls{IVA} service Pipeline.

\textbf{IVA and Big Dimensionality:} The \gls{VSDS} multi-modality can produce diverse types of data streams. Similarly, algorithms generate varied sorts of multi-dimensional features. The high-dimensionality factor poses many intrinsic challenges for data stream acquisition, transmission, learner, pattern recognition problems, indexing, and retrieval. In literature, it has been referred to as a “Big Dimensionality” challenge \cite{zhai2014emerging}. \gls{VSDS} variety leads to key challenges is how to acquire and process the heterogeneous data in an effective way. Most existing \gls{IVA} approaches can consider a specific input, but in many cases, for a single \gls{IVA} goal, different kinds, and formats can be considered. With growing features dimensionality, current algorithms quickly become computationally inflexible and, therefore, inapplicable in many real-time applications \cite{gao2017learning}. Dimension reduction approaches are still a hot research topic because of data diversity, increasing volume, and complexity. Effect-learning algorithms for first-order optimization, online learning, and paralleling computing will be more preferred.


\textbf{Model management:} An algorithm might hold a list of parameters. The model selection process encompasses feature engineering, \gls{IVA} algorithm selection, and hyperparameter tuning. Feature engineering is a strenuous activity and is influenced by many key factors, e.g., domain-specific regulations, time, accuracy, video data, and \gls{IVA} properties, which resultantly slow and hinder exploration. \gls{IVA} algorithm selection is the process of choosing a model that fixes the hypothesis space of prediction function explored for a given application \cite{friedman2001elements}. This process of \gls{IVA} algorithm selection is reliant on technical and non-technical aspects, which enforce the \gls{IVA} developer to try manifold techniques at the cost of time and cloud resources. Hyperparameter is vital as they govern the trade-offs between accuracy and performance. \gls{IVA} analysts usually do ad-hoc manual tuning by iteratively choosing a set of values or using heuristics such as grid search \cite{friedman2001elements}. From \gls{IVA} analysts’ perspective, model selection is an expensive job in terms of time and resources that bringing down the video analytics lifecycle. Model selection is an iterative and investigative process that generally creates an endless space, and it is challenging for \gls{IVA} analysts to know a priori which combination will produce acceptable accuracy/insights. In this direction, theoretical design trade-offs are presented by Arun et al. \cite{kumar2016model}, but further research is required that how to shape a unified framework that acts as a foundation for a novel class of \gls{IVA} analytics while building the procedure of model selection easier and quicker. 

\textbf{Statistics maintenance and ranking:} A user can develop and deploy an \gls{IVA} algorithm, model, or service that can be either extended, utilized, or subscribed by other users.  The community members run such architecture, and rapidly, the number of \gls{IVA} services can be reached to tons of domain-dependent or independent \gls{IVA} services. This scenario develops a complex situation for the users, i.e., which \gls{IVA} service (when sharing the parallel functionalities) in a specific situation, especially during service discover. Against each \gls{IVA} service, there is a list of \gls{QoS} parameters. Some of these \gls{QoS} parameters (not limited to) are user trust, satisfaction, domain relevance, security, usability, availability, reliability, documentation, latency, response time, resource utilization, accuracy, and precision. Such types of \gls{IVA} services against the \gls{QoS} parameters lead to the 0-1 knapsack issue. In this direction, one possible solution is utilizing multi-criteria decision-making approaches. It gives further opportunities to the research community to investigate how to rank and recommend \gls{IVA} algorithms, models, and services.

\textbf{IVAaaS and Cost Model:} The proposed \gls{SIAT} is supposed to provide \gls{IVAAaaS} and \gls{IVAaaS} in the cloud while adopting the \gls{C2C} business model. Unfortunately, current \gls{SaaS} cost models might not be applicable because of the involvement of diverse types of parameters that drastically affect the cost model. Such parameters are, business model (\gls{B2B}, \gls{B2C}, and \gls{C2C}), unite of video, user type (developer, researcher, and consumers), services (\gls{IVAAaaS} and \gls{IVAaaS}), service subscription (algorithm, \gls{IVA} service, single, multiple, dependent or independent), cloud resource utilization, user satisfaction, \gls{QoS}, location, service subscription duration, and cost model fairness. The addition of further parameters is subject to discussion, but the listed are the basic that govern \gls{SIAT} cost matrix. Additionally, the cost model demands further research and investigations to develop an effective price scheme for \gls{IVA} services while considering the stated parameters.

\textbf{Video big data management:} Despite video big data pose high value, but its management, indexing, retrieval, and mining are challenging because of its volume, velocity, and unstructuredness \cite{khan2020falkon}. In the context of video big data management, the main issue is the extraction of semantic concepts from primitive features. A general domain-independent framework is required that can extract semantic features, analyze and model the multiple semantics from the videos by using the primitive features. Further, semantic event detection is still an open research issue because of the semantic gap and the difficulty of modeling temporal and multi-modality features of video streams. The temporal information is significant in the video big data mining mainly, in pattern recognition. Limited research is available on content-based video retrieval while exploiting distributed computing. Further study is required to consider different types of features ranging from local to global spatiotemporal features utilizing and optimizing deep learning and distributed computing engines. Semantic-based approaches have been utilized for video retrieval because of the semantic gap between the low-level features and high-level human-understandable concepts \cite{alam2020intellibvr}. Ontology adds extra concepts that can improve the retrieval results and lead to unexpected deterioration of search results. In this context, a hybrid approach can be fruitful and need to design different query planes that can fulfill diverse queries in complex situations.

\textbf{Privacy, Security and Trust:} Video big data, acquisition, storage, and subscriptions to shared \gls{IVA} in the cloud become mandatory, which leads to privacy concerns. For the success of such platforms, privacy, security, and trust are always central. In literature, the word ‘trust’ is commonly used as a general term for ‘security’ and ‘privacy’. Trust is a social phenomenon where the user has expectations from the \gls{IVA} service provider and willing to take action (subscription) on the belief based on evidence that the expected behavior occurs \cite{khusro2017social}. In the cloud environment, security and privacy are playing an active role in the trust-building. To ensure security, the \gls{SIAT} should offer different levels of privacy control. The privacy and security phenomena are valid across \gls{VSDS}, storage security, multi-level access controls, and privacy-aware \gls{IVA} and analysis. 
\section{Conclusion}
\label{sec:conclusion}
In the recent past, the number of public surveillance cameras has increased significantly, and an enormous amount of visual data is produced at an alarming rate. Such large-scale video data pose the characteristics of big data. Video big data offer opportunities to the video surveillance industry and permits them to gain insights in almost real-time. The deployment of big data technologies such as Hadoop, Spark, etc., in the cloud under \gls{aaS} paradigm to acquire, persist, process and analyze a large amount of data has been in service from last few years.  This approach has changed the context of information technology and has turned the on-demand service model's assurances into reality. This paper presents a comprehensive layered architecture for intelligent video big data analytics in the cloud under the \gls{aaS}. Furthermore, research issues, opportunity, and challenges being raised by the uniqueness of the proposed \gls{CVAS}, and the triangular relation among video big data analytics, distributed computing technologies, and cloud has been reported.

\printglossary[type=\acronymtype, title= List of abbreviations]

%
%

\bibliographystyle{splncs04}
\bibliography{references}





\end{document}